\documentstyle[aps,12pt]{revtex}
\begin{document}
\author{S. T. R. Pinho, C. P. C. Prado, and O. Kinouchi}
\address{Instituto de F\'{\i}sica\\
Universidade de S\~{a}o Paulo\\
Caixa Postal 66318\\
05315-970, S\~ao Paulo, SP, Brazil\\
e-mail: prado@if.usp.br}
\date{September 23, 1997 }
\title{Absence of self-organized criticality in a random-neighbor version of the
OFC stick-slip model}
\maketitle

\begin{abstract}
We report some numerical simulations to investigate the existence of a
self-organized critical (SOC) state in a random-neighbor version of the OFC
model for a range of parameters corresponding to a non-conservative case. In
contrast to a recent work, we do not find any evidence of SOC. We use a more
realistic distribution of energy among sites to perform some analytical
calculations that agree with our numerical conclusions.

PACS number(s): , 64.60.L, 05.40, 05.70.L

\end{abstract}

The first cellular automaton to display SOC (self-organized criticality),
the sandpile model\cite{bak1}, required both a bulk conservation law and a
lattice structure with open boundary conditions, that were thought to be the
main ingredients to allow the building of the spatial correlations\cite
{kim2,Hwa}. There are some attempts to construct much simpler and more
tractable mean-field versions of these systems \cite{tang2,dhar3,janowsky4}.
Olami, Feder and Christensen\cite{OFC} introduced a non-conservative
cellular automaton (the earthquake model) that displays self-organized
critical behavior for a certain range of a parameter $\alpha $ related to
the breaking of a conservation law. A detailed analysis of this model\cite
{kim,grinstein} seems to show that the spatial inhomogeneities created by
the lattice, the particular boundary conditions, and the global driven
mechanism, are the essential ingredients for the existence of SOC. In a
recent publication, Lise and Jensen \cite{lise12} considered a
random-neighbor mean-field version of the OFC model with coordination $q=4$.
According to these calculations, it is remarkable that SOC can still be
observed in non-conservative cases, for $2/9=\alpha ^{*}<\alpha <\alpha
_c=1/4$.

In this work we revisit the random-neighbor version of the OFC model, with
an arbitrary coordination $q$ (that should be of interest in the context of
the investigations of SOC in artificial neural networks\cite{neurals}, in
which case the existence of a regular lattice is not a realistic
assumption). Unlike the conclusions of Lise and Jensen, we do not find any
evidence of SOC, except in the conservative case. We show (see figure 1)
that the use of lattices that are not big enough to allow the occurrence of
the largest avalanches can lead to the wrong impression that the system does
display SOC. Of course, this is an important word of caution about all
numerical evidences of SOC. Within the mean-field approach, we show that the
use of a simple and slightly more realistic assumption about the form of the
stationary probability distribution $p(E)$ of the energies of the sites is
enough to completely change the conclusions of Lise and Jensen. Using this
simplified form of $p(E)$, there is just a very small range of parameters
where SOC might occur in the non-conservative context.

The random-neighbor version of the OFC model consists of $N$ sites with an
energy $E_i<E_c$, for $i=1,...,N$, where $E_c$ is a threshold value (the
energies of the stable sites will be labelled by a superscript $-$). During
a long time scale, there is a continuous build up of all energies until the
energy of a certain site $i$ reaches the value $E_c$. Then, site $i$ becomes
unstable (its energy being labelled by a superscript $+$), and the system
relaxes in a very short time scale. The unstable site $i$ transfers an
energy $\alpha E_{+}$ $(\alpha \leq 1/q)$ to $q$ random neighbors, which may
also become unstable and may generate an avalanche that only stops when the
energies of all sites are again below $E_c$. The equality $\alpha =\alpha
_c=1/q$ corresponds to the conservative limit.

To make contact with the work of Lise and Jensen\cite{lise12}, we define a
''re-scaled'' dissipation constant, $\alpha _q=q\alpha /4$. According to the
same approximate calculations of Lise and Jensen, the critical value $\alpha
_q^{*}$ below which the avalanches have a typical size is given by $\alpha
_q^{*}=q/\left( 4q+2\right) $. If $q=4$, we have $\alpha _4^{*}=2/9=0.222$,
as obtained by Lise and Jensen.

Our own simulations, however, indicate that the density distribution of the
avalanche sizes $\rho (s)$ converges to a fixed curve that does not depend
on the size of the lattice, even for values of $\alpha _q$ far beyond $%
\alpha _q^{*}$, given by our mean-field calculations (see figures 1 and 2for 
$\rho \left( s\right) $ and the average avalanche size, $\left\langle
s\right\rangle $, repectively). Because of practical limitations, it is not
feasible to simulate systems closer to the conservative limit, but we
strongly believe that the collapse of $\rho (s)$ to a universal curve, and
the convergence of $\left\langle s\right\rangle $ to typical value,
regardless of the size of the lattice, for all values of $\alpha _q$ that we
were able to use in the simulations, do indicate that the system is not in a
critical state.

The results of numerical simulations for $p(E)$, with $q=4$, are shown in
figure 3 (for simplicity we assume from now on that $\alpha _4=\alpha $).
These results clearly show that $p(E)$ is not a simple constant. We then
decided to use the same approach of Lise and Jensen\cite{lise12}, but
supposing that the distribution $p(E)$ has the (more realistic) form shown
in the detail of figure 3, where $\Delta _p$ is half the width of each peak, 
$\Delta _b$ is the width of the gaps between two peaks, and $E^{*}=3\Delta
_b+7\Delta _p$ is the maximum value of $E$ for which $p(E)\neq 0$. We then
have 
\begin{equation}
\label{eq1}P_{+}(E^{+})=\frac{%
\displaystyle \int 
\limits_{E_c-\alpha E^{+}}^{E_c}\,p(E)\,\;dE}{%
\displaystyle \int 
\limits_0^\infty \;p(E)\;dE}=\frac 1{7a\Delta _p}%
\displaystyle \int 
\limits_{E_c-\alpha E^{+}}^{E_c}\,p(E)\,\;dE. 
\end{equation}
The lower limit of the integral in the numerator, $E_c-\alpha E^{+}$, can
belong to any of the intervals that define the four peaks of the
distribution, to which we assign the indices $i=1,2,3,4$. Now we have to
consider each one of these possibilities (to simplify the notation, we use
the superscript $i$ to refer to the value of $E_c-\alpha E_{+}$). The
integrals $P_{+}^i(E^{+})$ have the generic form 
\begin{equation}
\label{eq2}P_{+}^i(E^{+})=1+\frac{(\,i-1)\,\Delta _b}{7\Delta _p}-\frac{E_c}{%
7\Delta _p}+\frac{\alpha \,E^{+}}{7\Delta _p}. 
\end{equation}
Therefore, the branching rate $\sigma $ is given by 
\begin{equation}
\label{eq3}\sigma =4\,P_{+}^i=4\left[ 1+\frac{(\,i-1)\,\Delta _b}{7\Delta _p}%
-\frac{E_c}{7\Delta _p}+\frac{\alpha \left\langle E^{+}\right\rangle }{%
7\Delta _p}\right] . 
\end{equation}
Now, we write the mean-value of the energy of an unstable site, 
\begin{equation}
\label{eq4}\left\langle E^{+}\right\rangle ^i=\frac{\left\langle
E^{-}\right\rangle ^i}{(1-\alpha )}, 
\end{equation}
where 
\begin{equation}
\label{eq5}\left\langle E^{-}\right\rangle =\frac{%
\displaystyle \int 
\limits_{E_c-\alpha \left\langle E^{+}\right\rangle }^{E_c}E\;p(E)\;dE}{%
\displaystyle \int 
\limits_{E_c-\alpha \left\langle E^{+}\right\rangle }^{E_c}p(E)\;dE}. 
\end{equation}
Therefore, 
\begin{equation}
\label{eq6}\left\langle E^{+}\right\rangle ^i=\frac{E_c}{\alpha (2-\alpha )}-%
\frac{[7\Delta _p+(i-1)\Delta _b](1-\alpha )}{\alpha (2-\alpha )}\pm \frac{%
\sqrt{y_i}}{2\alpha (2-\alpha )}, 
\end{equation}
where 
\begin{equation}
\label{eq7}y_i=4\left\{ E_c\left( 1-\alpha \right) -\left[ 7\Delta _p+\left(
\,i-1\right) \,\Delta _b\right] \right\} ^2+4\alpha \,\left( 2-\alpha
\right) \,\left[ x_i-14\left( i-1\right) \right] \,\Delta _p\,\Delta _b, 
\end{equation}
with $x_i=24$, $26$, $32$, and $42$, for $i=1$, $2$, $3$, and $4$,
respectively.

To reach a critical state, we have to impose $\sigma \geq 1.$ Taking $\sigma
=1$, and using Eq. (\ref{eq6}), instead of Lise and Jensen's approximate
result, we have 
\begin{equation}
\label{eq8}7\Delta _p\left( 2+\alpha \right) +4\left( \,i-1\right) \Delta
_b-4E_c\left( 1-\alpha \right) +7\alpha E_c\pm 2\sqrt{\;y_i}=0. 
\end{equation}
For instance, if we take $\Delta _p=0.08$ and $\Delta _b=0.1$, the critical
branching condition leads to values of $\alpha ^{*}$ outside of the physical
range (that is, $\alpha ^{*}>1/4$). Therefore, in this particular case, it
is physically forbidden to assume that $\sigma =1$, so there is no
self-organized critical state.

In the conservative limit (for $\Delta _p\rightarrow 0$, $\Delta
_b\rightarrow \alpha E_c$), the four peaks of $p(E)$ tend to four delta
functions at $(\,i-1)\,\alpha \,E_c$. In this case, it is easy to see that $%
\sigma =1$ leads to the only possibility $\alpha ^{*}=\alpha _c=1/4$ (we
obtain $\alpha ^{*}>1/4$ for $i=1$, $2$, and $3$). It can also be shown
that, if we consider the limit $\Delta _b\rightarrow 0$ and $\Delta
_p\rightarrow E_c\,/\,7$, which corresponds to the approximation of Lise and
Jensen, then $\alpha ^{*}=2/9$.

In general, from Eq. (\ref{eq8}), for all values of $i$, the regions of
parameters associated with $\alpha \leq 1/4$ are determined by 
\begin{equation}
\label{eq9}E_c-\frac{175}{24}\Delta _p-\frac{2x_i}{21}\Delta _b\leq 0. 
\end{equation}
From this inequality, and the relation $E_c\geq 7\Delta _p+3\Delta _b$ (see
figure 3), we see that only in a very small region of the parameters $\Delta
_p$ and $\Delta _b$ (see figure 4) there are values of $\alpha ^{*}$ in the
physical range (that is, such that $0<\alpha ^{*}\leq 1/4$). In all of those
cases, $\Delta _b$ is very small, and the shape of $p(E)$ is very close to
the constant form used by Lise and Jensen.

In conclusion, on the basis of a mean-field argument, supplemented by a more
realistic approximation for the distribution of energies $p(E)$, we give
some analytical indications to support our own numerical findings that the
random-neighbor version of the OFC stick-slip model cannot display SOC
outside of the conservative limit. At the time we were writing this paper,
we came to know about two other works\cite{hakin,grass} that lead
essentially to the same conclusions.

Acknowledgments

We thank S. R. Salinas for helpful discussions. We acknowledge the financial
help of the brazilian organizations CAPES\ and FAPESP. One of us (STRP) is
on leave of absence from Instituto de F\'{\i}sica, Universidade Federal da
Bahia, Salvador, Bahia, Brazil.

\newpage\ 

Figure Captions

Fig. 1- a) Distribution of the avalanche sizes (number of topples) for $q=4$%
, $\alpha =0.23$, and different lattice sizes, $L$. The number of iterations
is $n=2\times 10^6$. For $L\geq 400$, the curves collapse to the same form,
which indicates that there is no self-organized critical state; b) Using $%
L=100$, $\alpha =0.23$, and $n=4\times 10^6$, the statistics is similar for
different coordinations, $q=4$, $6$, and $10$. However, for large $q$, and $%
\alpha $ not too large, the redistribution of energy is not sufficient to
generate larger avalanches.

Fig. 2- a) Average avalanche size $\left\langle s\right\rangle $ versus
lattice size $L$, for $q=4$, and different values of $\alpha _q$ (smaller
and larger than Lise and Jensen's value, $\alpha _q^{*}=2/9$). For $\alpha
_q^{*}=0.23$, for example, using bigger lattice sizes ($L=600$), we can see
the exponential behavior of $\left\langle s\right\rangle $ (in contrast to
Lise and Jensen's simulations). For $\alpha _q\geq 0.24$, it is necessary to
use much bigger lattice sizes to see this type of behavior; b) Average
avalanche sizes versus lattice size for $\alpha _q=0.23$, and different
coordination numbers, $q=2$, $4$, and $6$. For larger values of $q$, it is
sufficient to use a small lattice size to see the exponential behavior of $%
\left\langle s\right\rangle .$

Fig 3-a) Distribution of energy per site $p(E)$ versus energy $E$, for $q=4$%
, and $\alpha _q=0.21$, $0.22$, and $0.23$. The width of the peaks decreases
as $\alpha _q$ increases; b) Special form of $p(E)$ as used in the
calculations, where $\Delta _{p\text{ }}$ is the half-width of the peaks, $%
\Delta _b$ is the width of the gaps, $a$ is the value at the peak, and $%
E_c\geq E^{*}=7\Delta _p+3\Delta _b$.

Fig. 4-Space of parameters of $p(E)$ in terms of $\gamma _p=\Delta _p/E_c$
and $\gamma _b=\Delta _b/E_c$. The shaded regions correspond to the
intersection between $\alpha \leq 1/4$ and $\left( 7\Delta _p+3\Delta
_b\right) /E_c=7\gamma _p+3\gamma _b\leq 1$. Depending on the value of $%
E_c-\alpha E^{+}$, we have (a) $E_c-\alpha E^{+}$ $\epsilon $ $\left[
0,\Delta _p\right] $; (b) $E_c-\alpha E^{+}$ $\epsilon $ $\left[ \Delta
_p+\Delta _b,3\Delta _p+\Delta _b\right] $; (c) $E_c-\alpha E^{+}$ $\epsilon 
$ $\left[ 3\Delta _p+2\Delta _b,5\Delta _p+2\Delta _b\right] $; and (d) $%
E_c-\alpha E^{+}$ $\epsilon $ $\left[ 5\Delta _p+3\Delta _b,7\Delta
_p+3\Delta _b\right] $.

\end{document}